\documentclass[11pt]{article}
\usepackage{moriond_mypict,epsfig}

\def\Journal#1#2#3#4{{#1} {\bf #2}, #3 (#4)}

\def\NPB{{\em Nucl. Phys.} B}
\def\PLB{{\em Phys. Lett.}  B}
\def\PL{{\em Phys. Lett.} }

\def\PRD{{\em Phys. Rev.} D}
\def\ZPC{{\em Z. Phys.} C}
\def\EPJC{{\em Eur. Phys. J.} C}

\def\be{\begin{equation}}
\def\ee{\end{equation}}
\def\bea{\begin{eqnarray}}
\def\eea{\end{eqnarray}}

\begin{document}
\vspace*{4cm}
\title{RESULTS ON SEARCHES BEYOND THE STANDARD MODEL AT HERA}

\author{ D. BOSCHERINI }

\address{I.N.F.N., Department of Physics, Via Irnerio 46,\\
40126 Bologna, Italy\\
(on behalf of the H1 and ZEUS collaborations)}

\maketitle\abstracts{
Signals of physics beyond the Standard Model at HERA
have been searched for by the H1 and ZEUS collaborations.
Results are reported about searches of contact interaction effects,
leptoquarks, R-parity violating squarks, excited fermions and
single top production.
The analyses used the data set collected during the HERA phase I,
consisting in about 110 (15) pb$^{-1}$ of {\it e$^+$p}
({\it e$^-$p}) collisions per experiment.
No evidence for new physics has been found and limits have been set
on the parameters of the models considered.
}

\section{Introduction}
At HERA electrons or positrons of 27.5 GeV and protons of 920 GeV
(820 GeV before 1998) are collided, resulting in a center-of-mass energy,
$\sqrt{s}$, of 318 (300) GeV.
The maximum square momentum transfer ($Q^2$) reached is a few times
$10^4$ GeV$^2$, which means the resolution power in probing the proton
structure is of the order of 10$^{-16}$cm.

In the years 1994-2000 the luminosity collected by each experiment
is approximately 110 pb$^{-1}$ with $e^+p$ and 15 pb$^{-1}$
with $e^-p$ collisions.

Signals for new physics have been searched by both collaborations 
and are reported for the following topics:
\begin{itemize}
\item
Physics at an energy scale much higher than the HERA energy
could produce deviations in the high $Q^2$ region.
Modeling the process as a four fermion contact interaction ({\it eeqq})
the evidence for new physics can be investigated.
\item
Resonant electron-quark states can be formed, taking advantage of
the unique opportunity of an initial state with definite leptonic and
baryonic number provided by HERA.
Leptoquarks and R-parity violating squarks are examples of such states.
\item
Excited states of leptons or quarks, predicted by compositeness theories,
could be produced if their masses are below the HERA center-of-mass energy.
\item
Events with a high $p_T$ lepton and missing transverse energy have been
searched, following the first observation by H1.
Single top production events at HERA would present the same topology
and could be observed if an anomalous large top coupling existed.
\end{itemize}

\section{Contact Interactions}
Four-fermion contact interactions (CI) are an effective theory which can be
used to describe virtual effects coming from physics processes at
much higher energy scale.
Such effects could produce deviations on the observed $Q^2$ distributions
in Deep Inelastic Scattering (DIS) with respect to Standard Model (SM)
expectations, which could result in an increase of the
cross section at the largest $Q^2$ and an interference effect
(positive or negative) with the SM in the intermediate $Q^2$ region
\cite{ci-effects}.

As strong limits have already been placed on the scalar and tensor
terms \cite{ci-vec-tens-lim}, only the vector {\it eeqq} terms have been
considered here. They can be represented as additional terms in the
SM Lagrangian:
\begin{equation}
\label{eq:ci}
{L}_{CI} = \sum_{\alpha,\beta=L,R}^{q=u,d} \; \eta_{\alpha\beta}^q ( \bar {e}_\alpha \gamma^\mu e_\alpha) (\bar q_\beta \gamma_\mu q_\beta)
\end{equation}
where the coefficients $\eta_{\alpha\beta}^q = \epsilon g_{CI}^2 / \Lambda^2$
characterize the different CI scenarios defining their chiral properties
via the sign of the interference with SM, $\epsilon$,
their strength via the coupling $g_{CI}$
(by convention set to $g_{CI} = \sqrt{4\pi}$)
and their effective mass scale via the parameter $\Lambda$
($\alpha$ and $\beta$ are, respectively, the electron and quark helicity
and $q$ is the quark flavor).

\begin{figure}
\centerline{
\hbox{\psfig{figure=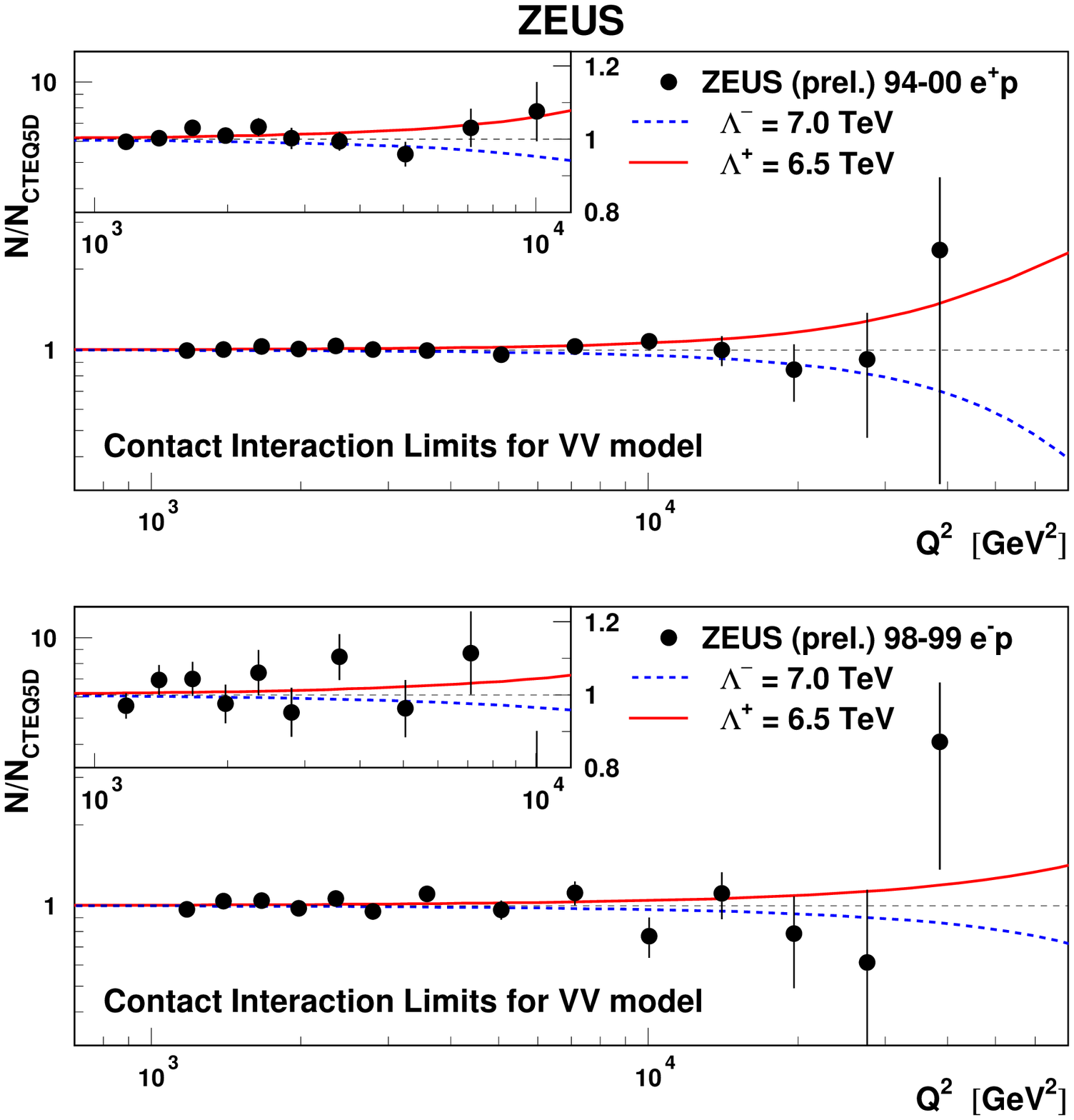,height=3.1in}}
\hbox{$\;\;\;\;\;$}
\hbox{\psfig{figure=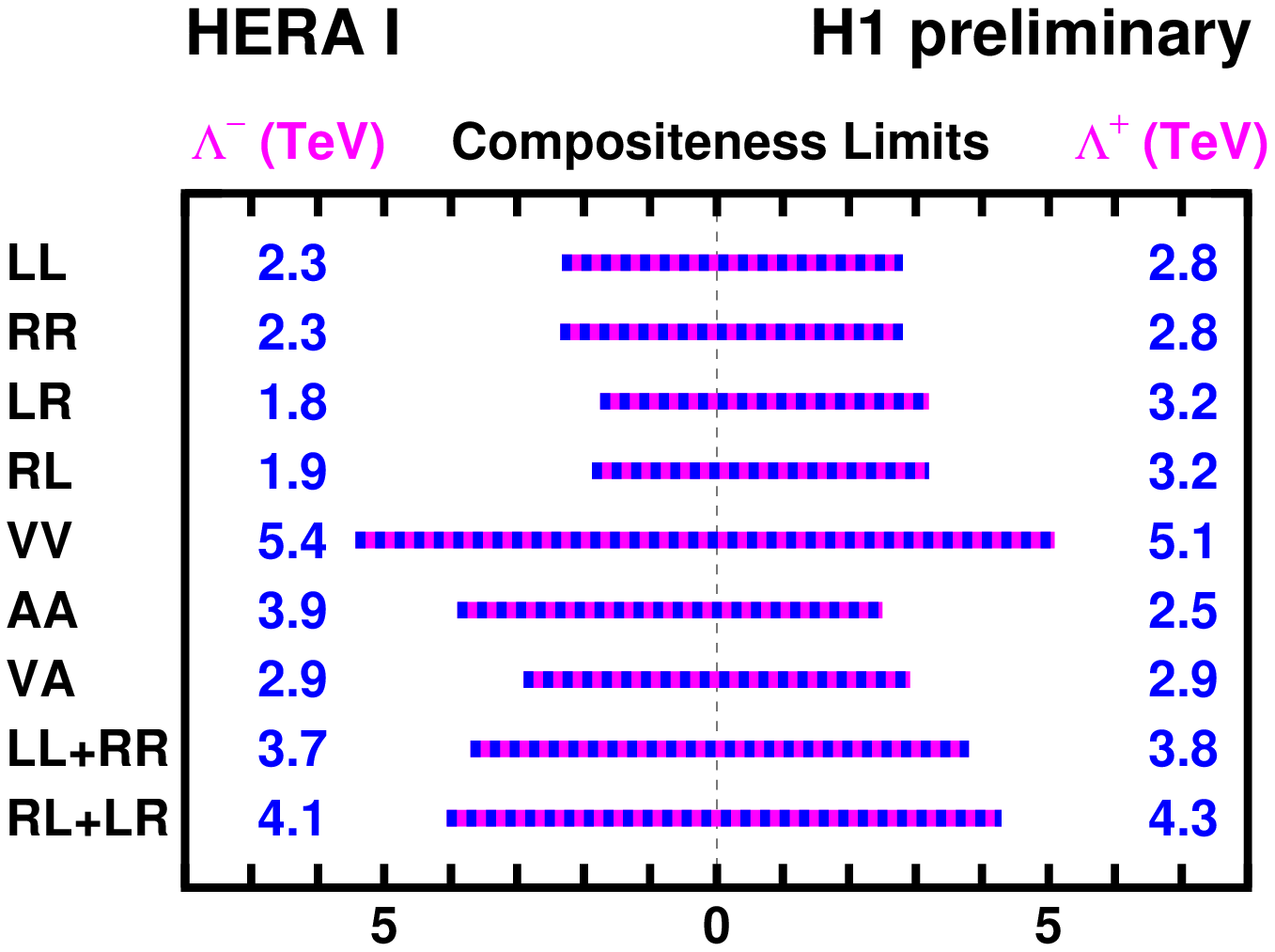,height=2.2in}}
}
\caption{Left side:
ZEUS Neutral Current measurements (normalized to SM expectations using the CTEQ5D
parton distributions) compared with curves corresponding to 95\% CL exclusion limits
for the effective mass scale in the VV contact-interaction model, for positive
($\Lambda^+$) and negative ($\Lambda^-$) couplings.
Right-side:
H1 limits on the compositeness scale $\Lambda$ resulting from the
combined analysis of the $e^+p$ and $e^-p$ data (values outside the regions delimited by the bars are excluded at 95\% CL). 
\label{fig:h1-zeus-ci}}
\end{figure}

The high $Q^2$ Neutral Current (NC) DIS measurements performed by both
collaborations~\cite{h1-ci,zeus-ci} show no deviations from the SM expectation,
hence limits on the scale $\Lambda$ were derived for the most common CI models.
The $Q^2$ distribution measured by ZEUS is shown in fig.~\ref{fig:h1-zeus-ci}
for the $e^+p$ (left top) and $e^-p$ (left bottom) data together with
the 95\% CL effect of the presence of a VV-type CI scenario
($\eta_{LL}=\eta_{LR}=\eta_{RL}=\eta_{RR}$).
The recently released limits obtained by H1 for several scenarios are shown
in fig.~\ref{fig:h1-zeus-ci}, right plot.
The limits range up to more than 5 TeV and are comparable to what obtained
at LEP and Tevatron studying reactions complementary to HERA, respectively
$e^+e^- \rightarrow q\bar q$ and Drell-Yan processes.

The specific model with Large Extra Dimensions~\cite{lxd} has also been
considered. In such a model SM particles can
propagate in the ordinary 4-dimensional space and the graviton
has access to the extra dimensions which are compactified to a size
below the millimeter scale.
With very large extra dimensions, the effective Planck scale, $M_S$, could
be comparable to the electroweak scale and CI-like terms with coupling
coefficient $\eta_G = \lambda / M_S^4$
could be visible at HERA in the high $Q^2$ region~\cite{grw}.
Limits on the effective scale $M_S$ have been set to $\sim$ 0.8 TeV
at 95\% CL~\cite{h1-lxd,zeus-lxd}.

\section{Leptoquarks}
Leptoquarks (LQ) are color-triplet bosons carrying both leptonic (L)
and baryonic (B) number predicted by several extensions of the SM.
They could be directly produced at HERA by fusion of the initial state
electron with the quark of the incoming proton with a cross section depending
on the unknown Yukawa coupling $\lambda$ at the LQ-electron-quark vertex.

The experimental results have been interpreted in the
framework of the Buchm\"uller-R\"uckl-Wyler (BRW) model~\cite{brw},
where the branching ratios
$\beta_e$ ($\beta_\nu$) for the LQ decays to $eq$ ($\nu q$) 
are assumed to have fixed values (1, 1/2, 0).
The BRW model classifies 14 possible scalar or vector LQ species with
fermionic number $F = 3B + L = 0$ or $|F|=2$.
The $F=0$ LQs couple to a particle and an anti-particle and are therefore
better tested with $e^+p$ collisions, while $|F| = 2$ LQs couple to
two particles or two anti-particles and are better tested
in $e^-p$ collisions.

The LQ decay to lepton-quark pair produces a final state identical to
SM DIS, but the angular distribution of the scattered lepton is different.
The inelasticity variable, $y$, is related to the lepton decay angle
in the lepton-quark rest frame, $\theta^*$, with $cos \theta^* = 1-2y$.
DIS events have a $1/y^2$ angular distribution, scalar LQs are
flat in $y$ and vector LQs have a $(1-y^2)$ dependence.
Therefore a cut in $y$ is used to enhance the LQ signal over the
DIS background.

\begin{figure}
\centerline{
\psfig{figure=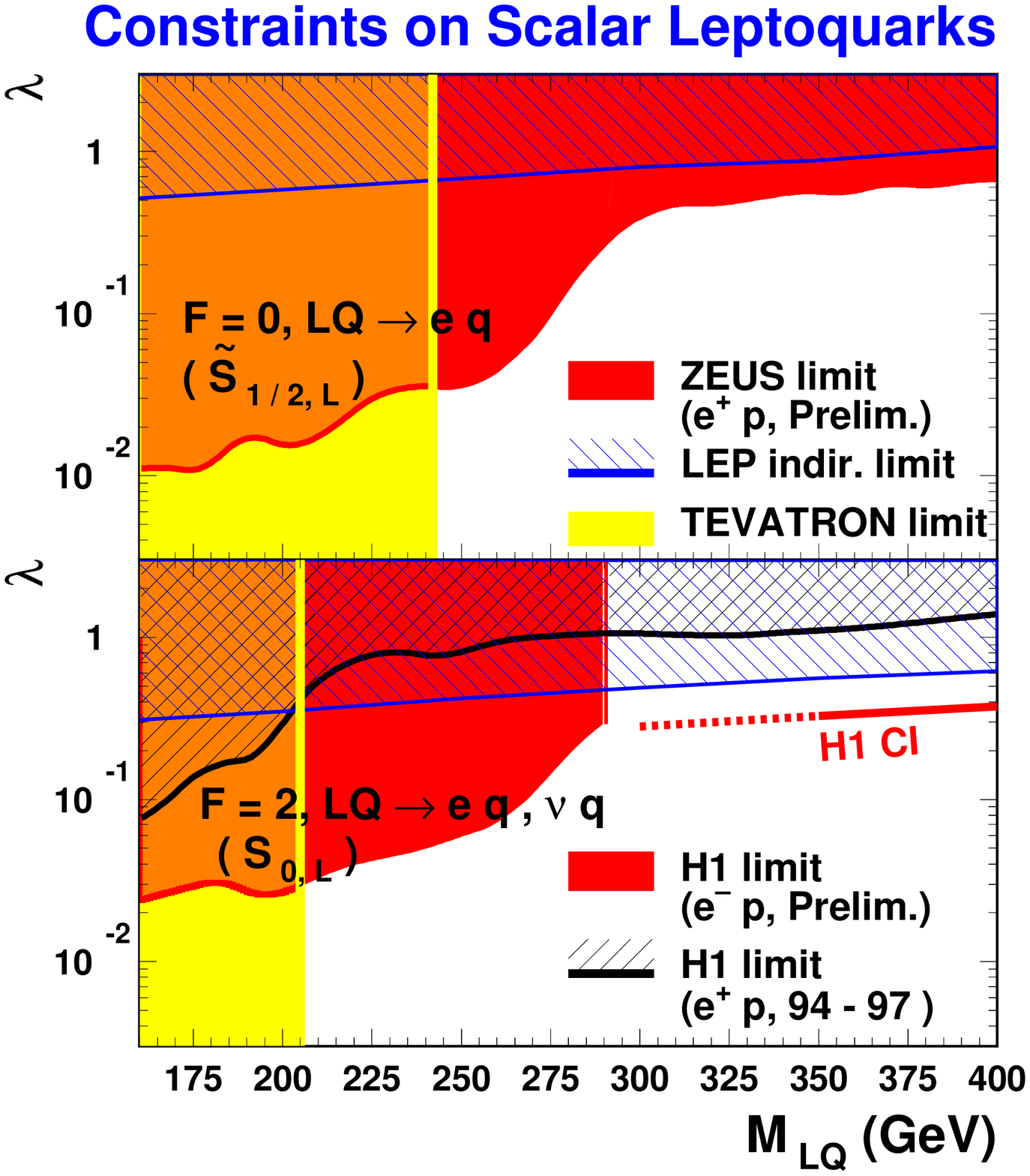,height=3.3in}
\hbox{$\;\;\;\;\;\;\;\;\;\;$}
\psfig{figure=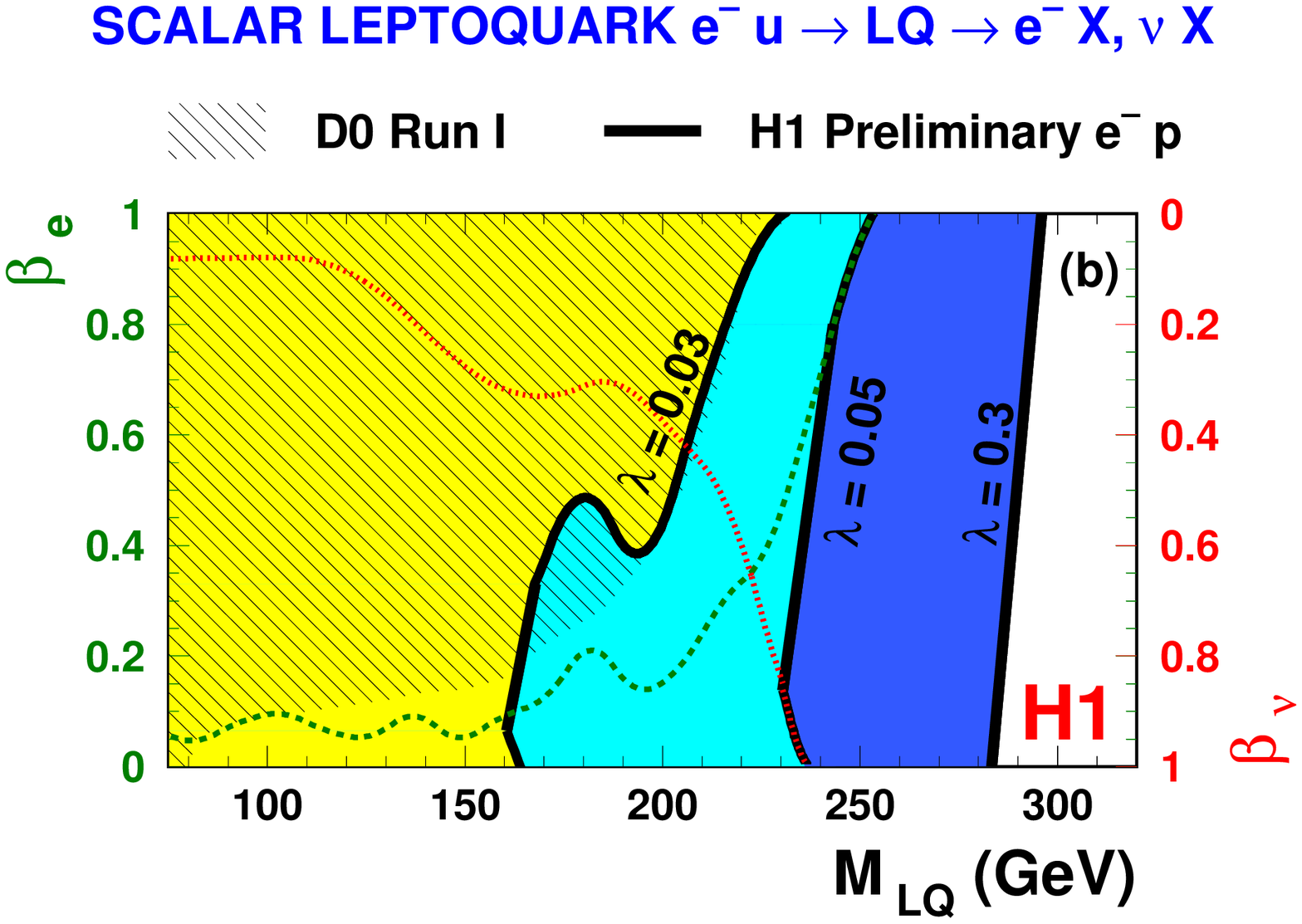,height=2.2in}
$\put(-280,175){\bf (a)}$
}
\caption{a):
Exclusion limits at 95\% CL on the Yukawa coupling $\lambda$ as
a function of the mass for a scalar LQ with $F=0$ in ZEUS (upper)
and with $F=2$ in H1 (lower).
The regions above the curves are excluded. LEP and Tevatron limits are also
shown.~~~
b):
Domains excluded by the combinations of NC
and CC data for a scalar $F=0$ LQ decaying only in $eq$ or $\nu q$ for
three values of the Yukawa coupling $\lambda$. The regions on the
left of the curves are excluded at 95\% CL.
For $\lambda = 0.05$ the part of the plane to the left of the dashed (dotted)
curve is excluded by the NC (CC) analysis.
Tevatron exclusion region is also shown.
\label{fig:lq-coupling-limits}}
\end{figure}

No evidence of a signal has been found by either collaboration
~\cite{h1-lq,zeus-lq},
therefore limits on the Yukawa
coupling as a function of the LQ mass have been derived and shown in
fig.~\ref{fig:lq-coupling-limits} (a) for two types of scalar LQs,
together with the LEP and Tevatron limits.
A more general LQ model has been considered, treating the branching
ratios $\beta_e$ and $\beta_\nu$ as free parameters, with the only assumption
that $\beta_e + \beta_\nu = 1$.
In this case, combining NC and CC results, the limits obtained are largely
independents of the branching ratios and are shown in
fig.~\ref{fig:lq-coupling-limits} (b).
The comparison with the Tevatron limits shows that HERA results extend
to lower masses and smaller branching ratio to $\nu q$.

\section{R-parity violating squarks}
R-parity is a multiplicative quantum number defined as $R_p=(-1)^{L+3B+2S}$
where L, B and S denote leptonic number, baryonic number and particle spin.
It distinguishes standard particles having $R_p = +1$,
from their supersymmetric partners for which $R_p = -1$.

In supersymmetric models with R-parity conservation
(e.g. MSSM, the Minimal Supersymmetric Standard Model), sparticles are
produced in pairs and the lightest supersymmetric particle is stable.
In models where R-parity is violated \cite{rp-violating-susy}, sparticles can
be singly produced and decay back to SM particles.

The $\lambda^\prime_{ijk}L_iQ_jD_k$ term in the supersymmetric lagrangian
is of interest at HERA
(L and Q are the left-handed lepton and quark doublet superfields,
D is a right-handed singlet superfield of down-type quarks, $\lambda^\prime$ is
the Yukawa coupling and $i$, $j$, $k$ denote their generation),
since a non-zero value of $\lambda^\prime_{1jk}$ would imply
the possibility to produce single squarks via $eq$ fusion.

\begin{figure}
\begin{center}
\psfig{figure=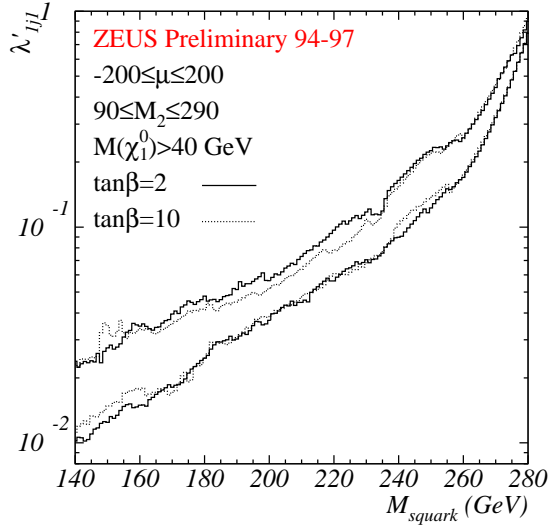,height=3.0in}
\end{center}
\caption{Minimum and maximum 95\% CL limits on $\lambda^\prime_{1j1}$
versus the squark mass for $\tan \beta = 2$ (solid) and 10 (dotted),
obtained in the region -200 GeV $\le \mu \le$ 200 GeV and 90 GeV
$\le M_2 \le$ 290 GeV.
\label{fig:rp-susy}}
\end{figure}

The decay mechanism of a squark can then involve R-parity conserving
decay to gauginos and produce many final states, made of
multi-jet and leptons, which have been investigated by both
collaborations~\cite{h1-rp-violation,zeus-rp-violation}.
No evidence of a signal has been found and in fig.~\ref{fig:rp-susy}
limits on the coupling as a function of the squark mass are shown
for two sets of SUSY parameters.

\section{Excited Fermions}
The family structure and mass hierarchy of the known fermions
suggests they could not be elementary particles but have a substructure,
as postulated in compositeness models.
A consequence of these models is the existence of excited states of leptons
and quarks which could be produced at HERA, depending on the dynamics of the
unknown sub-constituents.

Once produced, an excited fermion would decay to a normal fermion
by radiating off a gauge boson.

The experimental results have been studied in the theoretical framework of
the Hagiwara, Komamiya and Zeppenfeld model~\cite{hkz}, where the production
cross section depends on the relative coupling strengths $f$, $f^\prime$
and $f_s$ to the SU(2), U(1) and SU(3) gauge groups.

Assuming relations between the couplings, the cross section depends
only on the parameter $f/\Lambda$. The conventional assumptions used at
HERA are: $f_s = 0$ and $f = \pm f^\prime$.

\begin{figure}
\centerline{
\hbox{\psfig{figure=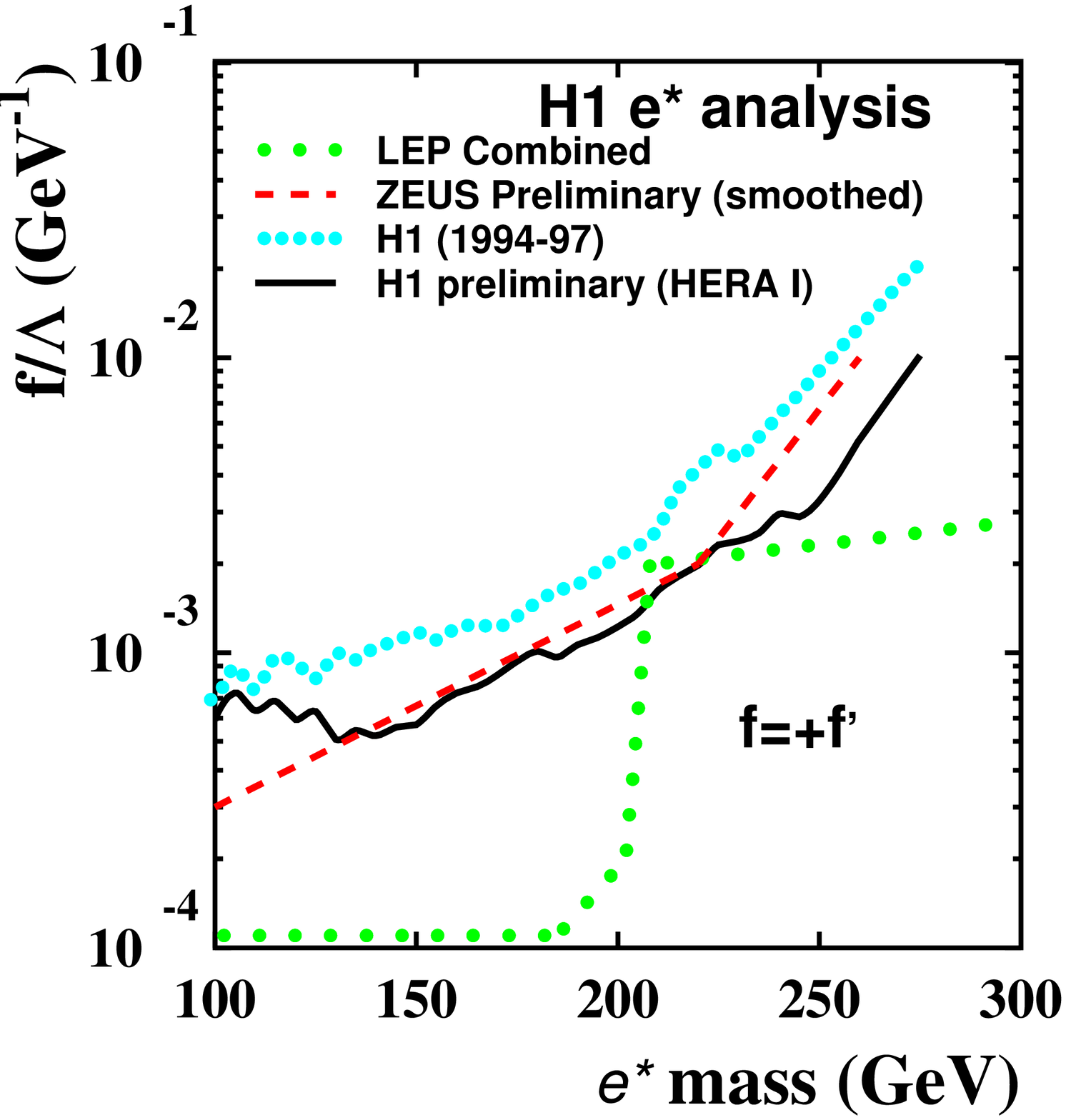,height=3.0in}}
\hbox{$\;\;\;\;\;\;$}
\hbox{\psfig{figure=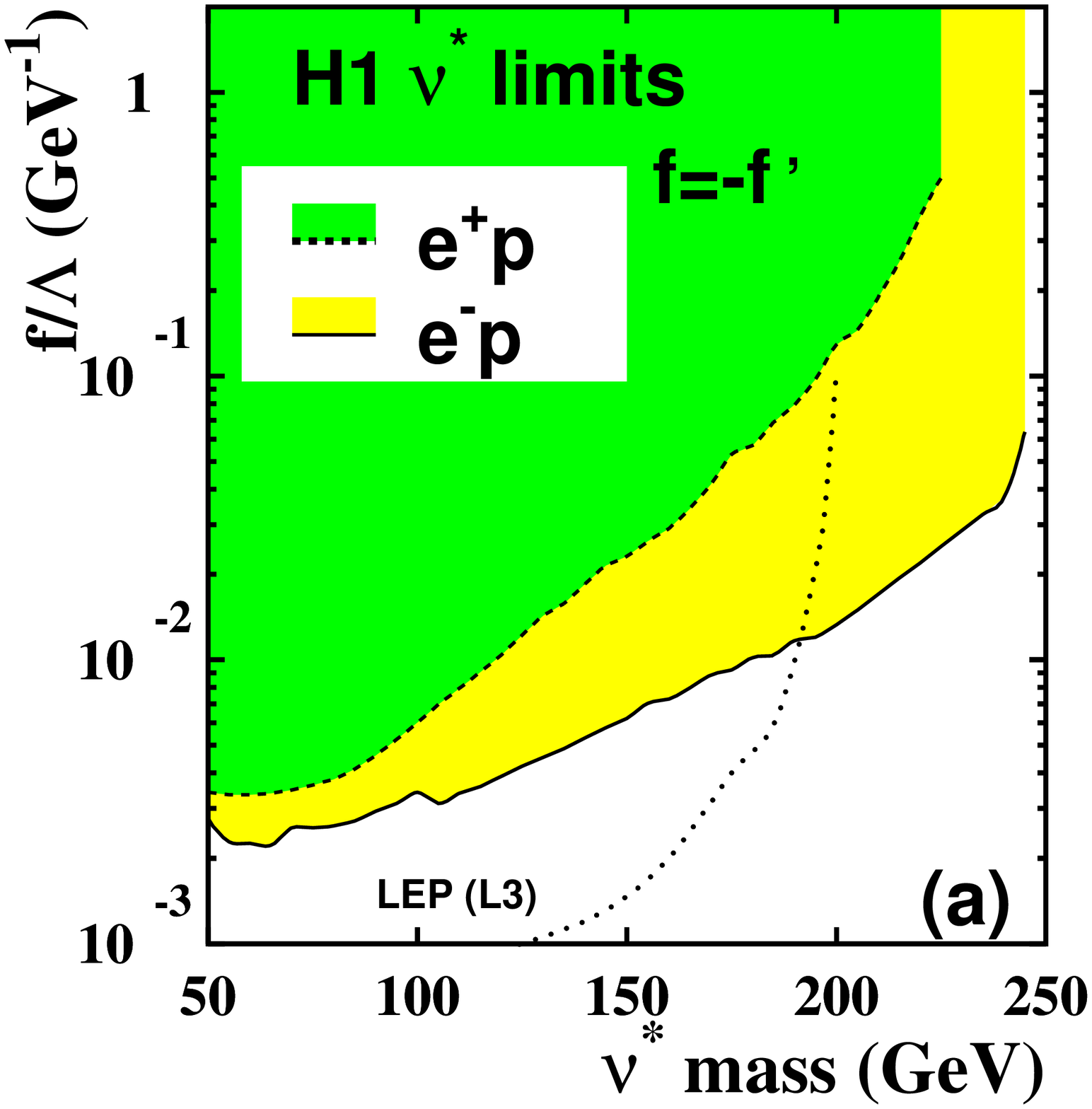,height=3.0in}}
}
\caption{Exclusion limits on the coupling $f/\Lambda$ at 95\% CL
as a function of the mass of the excited electron (left)
and excited neutrino (right).
LEP limits are also shown.
\label{fig:h1-fstar-limits}}
\end{figure}

Neither collaboration have found deviations from SM expectation
in any of the decay channels considered~\cite{h1-fstar,zeus-fstar}
and therefore limits have been set on $f/\Lambda$.
In fig.~\ref{fig:h1-fstar-limits} the H1 limits for excited electrons
and excited neutrinos are shown.
The excited electron limits are new preliminary results
obtained by considering all the electroweak decays
$e^* \rightarrow e + \gamma$, $e^* \rightarrow eZ$, $e^* \rightarrow \nu W$
and final states resulting from $Z$ or $W$ hadronic decays, and using
the full 1994-2000 luminosity.
The comparison to the LEP results shows that the HERA potentiality
is especially good in the excited neutrino search, where the limits
have been extended to masses above 200 GeV.

\section{Isolated Lepton Events}
In 1998 H1 reported the observation of an excess in the production of
events with a high transverse momentum lepton and missing $P_T$ \cite{h1-iso-old}.
Both collaborations~\cite{h1-iso,zeus-iso-and-top} have analysed the complete data set available
from HERA phase I, and the results are summarized in
table~\ref{tab:iso-leptons}, where the data are compared with the SM
expectation (dominated by the $W$ production whose cross section has
been measured at HERA to be $\sim$1 pb~~\cite{w-xsec}).

It is clear that while the ZEUS result is in agreement with the SM expectations
both in the electron and in the muon channel, H1 observes an
excess of events.
On the other hand the MC expectations are in good agreement
between the two experiments.
\begin{table}[t]
\caption{Results of the search for isolated lepton events as observed
by H1 and ZEUS for two values of the hadronic transverse momentum
($P_T^X$) cut. The H1 value in the table refers to the $e^+p$ data only 
(see text for the H1 $e^-p$ result).
\label{tab:iso-leptons}}
\vspace{0.4cm}
\begin{center}
\begin{tabular}{|c|c|c|c|}
\hline
  & $P_T^X$ cut &        Electrons            &          Muons            \\
  &             & \tiny Observed/Expected (W) & \tiny Observed/Expected (W) \\
\hline
{\bf ZEUS} prel.  & 25 & $1 / 1.14\pm 0.06$ (1.10) & $1 / 1.29\pm0.16$ (0.95)\\
\tiny 94-00 $e^{\pm}$p 130 pb$^{-1}$ & 40 &  $0 / 0.46\pm 0.03$ (0.46) & $0 / 0.50\pm0.08$ (0.41) \\
\hline
{\bf H1} prel. & 25 & $4 / 1.29\pm 0.33$ (1.05) & $6 / 1.54\pm0.41$ (1.29)  \\
\tiny 94-00 $e^+$p  102 pb$^{-1}$  & 40  &  $2 /  0.41\pm 0.12$ (0.40) & $4 / 0.58\pm0.16$ (0.53)  \\
\hline
\end{tabular}
\end{center}
\end{table}

The H1 result for the 98-99 $e^-p$ data (14 pb$^{-1}$), not reported in the
table, is in agreement with the SM (Observed/Expected = 0/1.8$\pm$0.4).

\begin{figure}
\centerline{
\psfig{file=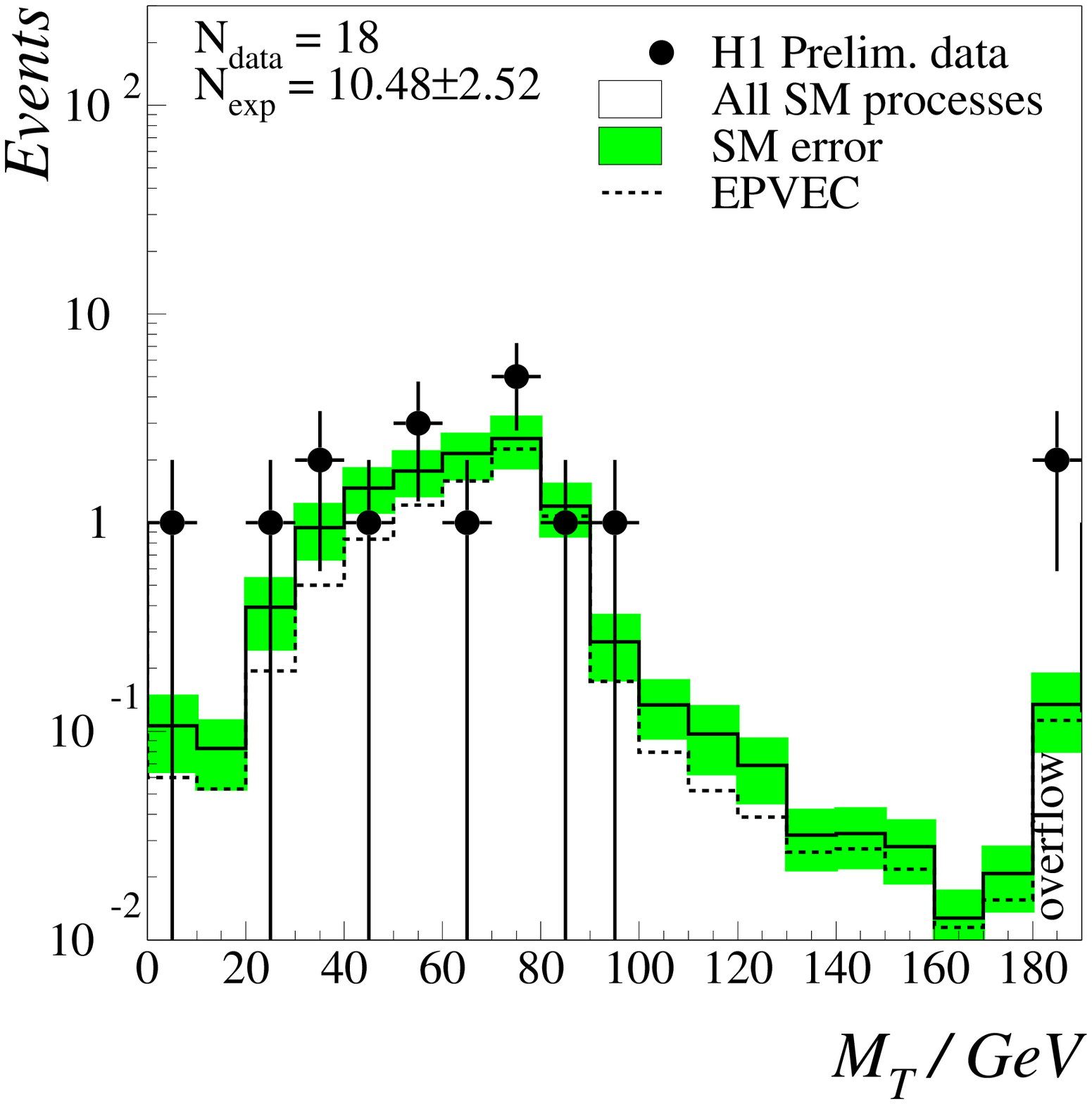,height=3.0in}
\hbox{$\;\;\;\;\;\;\;\;\;$}
\psfig{figure=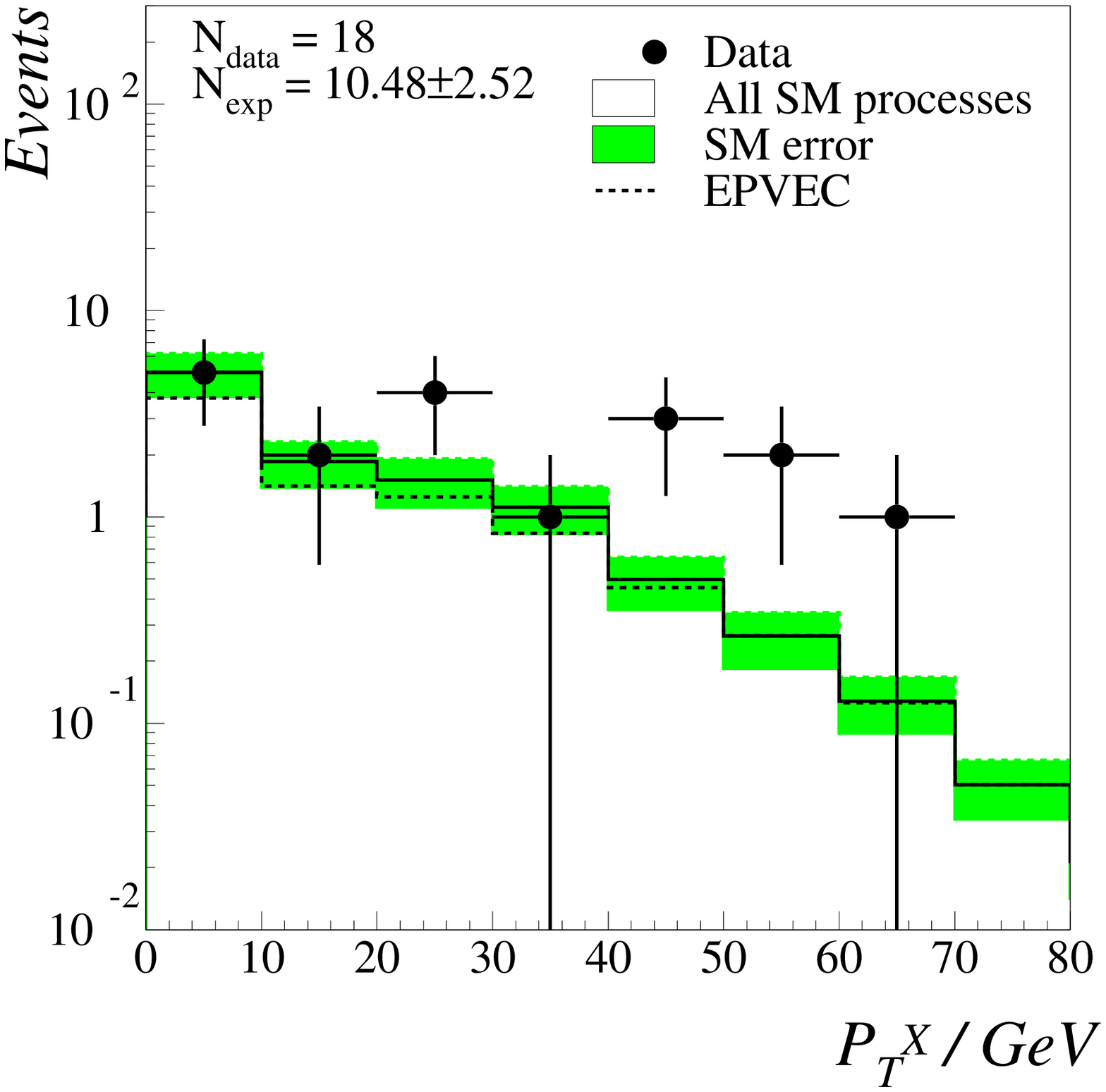,height=3.0in}
}
\caption{Transverse mass (left) and transverse hadronic momentum (right)
distributions for the final $e^+p$ data
selection in the combined electron and muon channels as measured in H1.
The total expectation from Standard Model is shown as full histogram together
with the total error band. The dashed histogram represents the $W$ production
component.
\label{fig:h1-mt-and-ptx}}
\end{figure}

The invariant mass distribution of the events is compatible
with $W$ production (fig.~\ref{fig:h1-mt-and-ptx} left).
The excess of events is pronounced in the kinematic region at high
hadronic transverse momentum, as shown in fig.~\ref{fig:h1-mt-and-ptx} right.

\section{Single Top Production}
The signature of the events discussed in the previous section is also
typical of a top quark decaying via $t \rightarrow bW \rightarrow bl\nu$.
Top production at HERA would proceed via Flavor Changing Neutral Current
(FCNC) events and is therefore tiny ($\sigma ( ep \rightarrow etX ) \sim 1$fb
at $\sqrt{s} = 300$ GeV) in the SM,
because FCNC are forbidden at tree level and the contribution from
loop diagrams is GIM suppressed.
However SM extensions predict anomalously large top couplings which
could increase the top production cross section to a sizable value.

For the leptonic decay of the $W$, H1~\cite{h1-top} observes five events
(3 electrons + 2 muons)
compatible with single top production, with an excess over the SM expectation
(Observed/Expected = 5/1.8). ZEUS~\cite{zeus-iso-and-top}, instead, is
in agreement with SM
(Observed/Expected = 0/0.96).

\begin{figure}
\centerline{
\psfig{figure=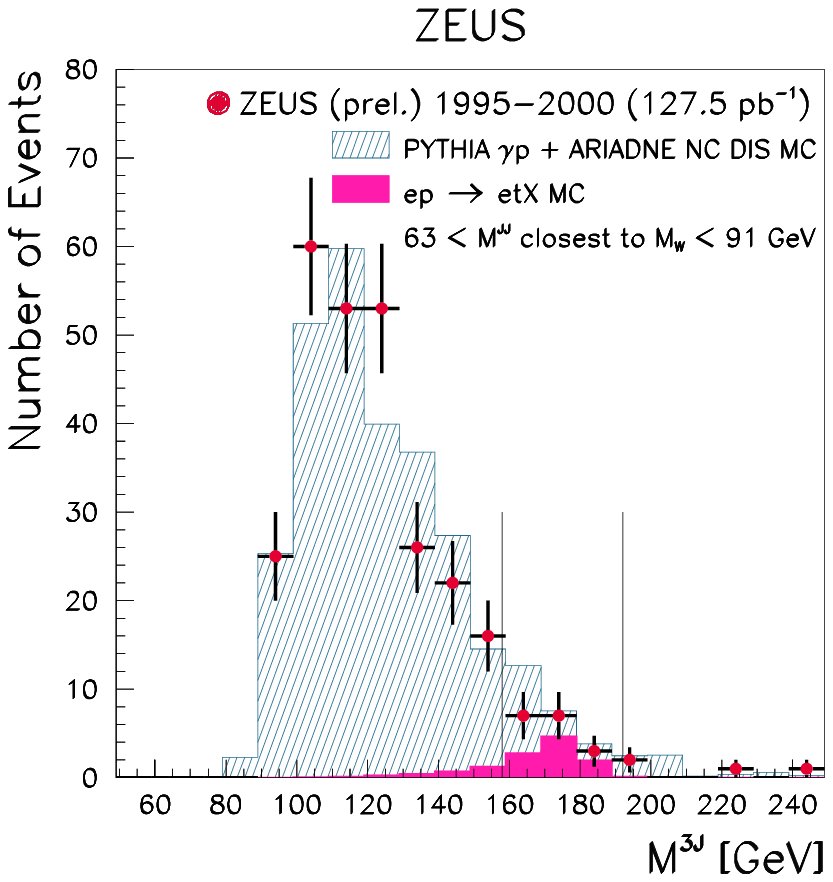,height=3.0in}
\hbox{$\;\;\;\;\;\;\;\;\;$}
\psfig{figure=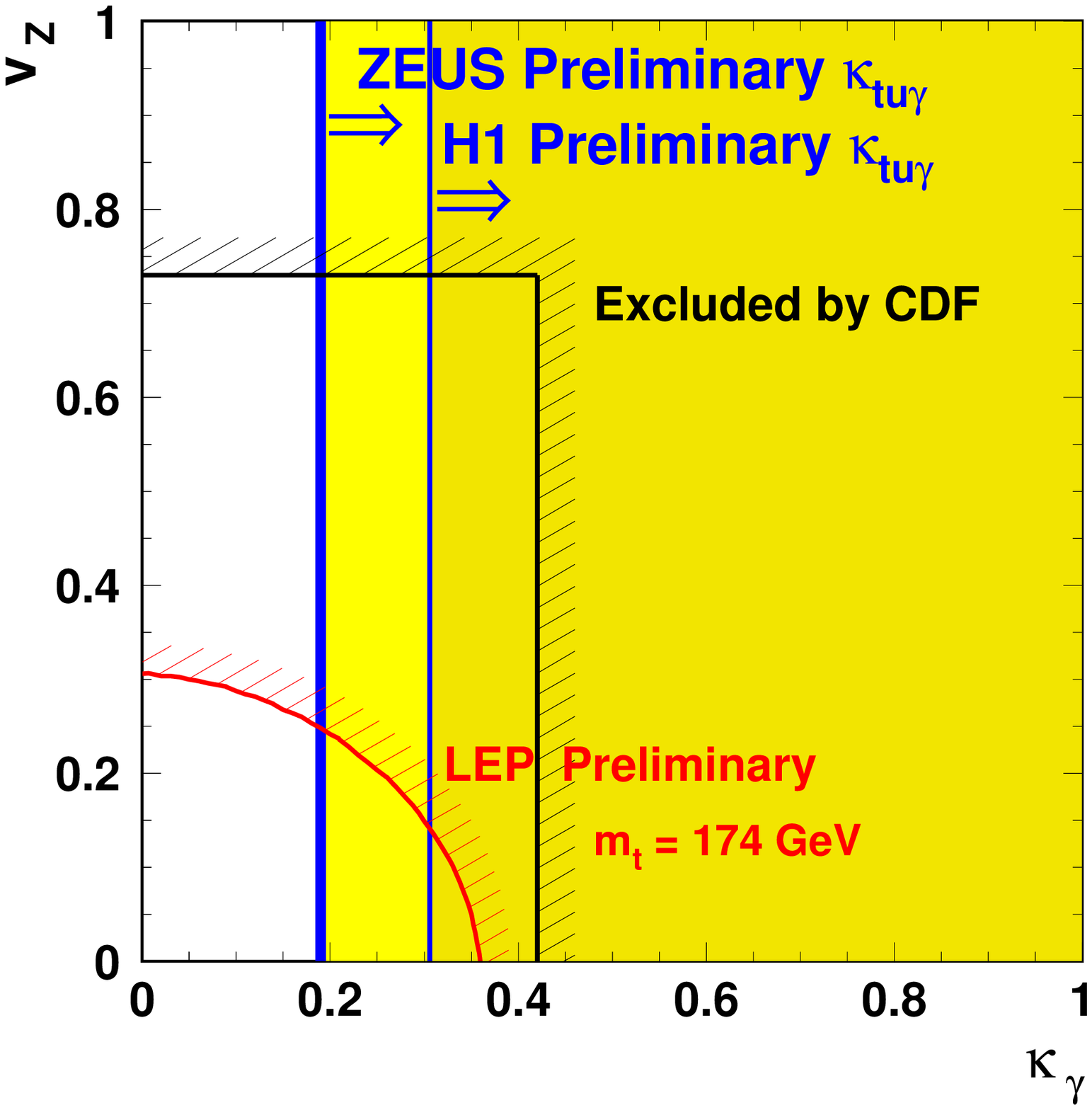,height=3.2in}
}
\caption{Left: three-jet invariant mass as reconstructed by ZEUS in the
single top selection in the hadronic channel. 
The mass cuts applied in the final selection are shown as dashed lines.~~~
Right: limits at 95\% CL on the FCNC magnetic coupling at the photon
vertex, $\kappa_\gamma$ and the vector coupling at the $Z$ vertex, $V_Z$
The LEP and Tevatron results are also shown.
HERA results apply only to $\kappa_{tu\gamma}$.
\label{fig:top-mass-and-limits}}
\end{figure}

The search in the hadronic decay channel has produced results consistent
with the SM for both collaborations and the ZEUS invariant mass plot
is shown in fig.~\ref{fig:top-mass-and-limits} left for the three-jet event candidates.
It has to be noticed that the sensitivity of the hadronic channel
is lower with respect to the leptonic one. This means the H1 excess
in the leptonic channel is not in contradiction with the hadronic channel
result.

Combining leptonic and hadronic results a limit has been set
on the magnetic coupling at the photon-top-quark vertex, which is
plotted in fig.~\ref{fig:top-mass-and-limits} right against the vector
coupling at the $Z$-top-quark vertex.

The HERA results, plotted together with the Tevatron and LEP results,
are absolutely competitive.
The difference between ZEUS and H1 is simply due to the events found
by H1 in the final selection.

\section{Summary}
The data set collected during HERA phase I has been extensively analysed
by the H1 and ZEUS collaborations in the search for new physics.
No evidence for new processes beyond the SM has been found.
The exclusion power at HERA is similar to that at the other
colliders (LEP and Tevatron).

The luminosity of the HERA II programme is needed to clarify the
still outstanding isolated lepton events found by H1.
In contrast to that, ZEUS data are in good agreement with SM.

The HERA II programme has reached its startup phase.
The promised increase in luminosity should add up to
an integrated value of about 1 fb$^{-1}$ per experiment
by the end of 2006, a factor 10 higher than the current value.
This will allow an improvement of the limits currently set for
contact-interactions, leptoquarks, squarks and excited fermions.

The running options with polarized $e^\pm$ beams will allow to
enhance and/or disentangle the processes coupling to chiral leptons
and to better control some background process.

Considerable efforts also on the detector side have been undertaken.
Both experiments are now equipped with an improved forward tracking and
vertex reconstruction capability, which will increase the sensitivity
to flavor specific processes.


\end{document}